# M2M Communications for E-Health and Smart Grid: An Industry and Standard Perspective

Zhong Fan, Russell J. Haines, and Parag Kulkarni
Telecommunications Research Laboratory, Toshiba Research Europe Ltd.,
32 Queen Square, Bristol, BS1 4ND, UK
{zhong.fan, russell.haines, parag.kulkarni}@toshiba-trel.com

*Abstract*—An overview of several standardization activities for machine-to-machine (M2M) communications is presented, analyzing some of the enabling technologies and applications of M2M in industry sectors such as Smart Grid and e-Health. This summary and overview of the ongoing work in M2M from the industrial and standardization perspective complements the prevalent academic perspective of such publications to date in this field.

## I. INTRODUCTION

Machine-to-Machine (M2M) communication has attracted a lot of attention in both academia and industry recently [1], finding many applications including Smart Grid, Smart Community, consumer electronics, vehicular communications, industrial control and automation (SCADA - Supervisory Control And Data Acquisition), point-of-sale/retail/ATMs and e-Health. M2M promises to increase network connections by at least an order of magnitude – a study by analysts Berg Insight predicts a growth from 47.7 million cellular connections in 2008 to 187 million by 2014, a compound annual growth rate (CAGR) of 25.6% [2], driven by M2M devices. This is a vast and rapidly growing area.

M2M communications, a key enabling technology for the Internet of Things (IoT), allows different devices (e.g., smart phones, game consoles, computers, cameras, cars, sensors, personal health devices) to communicate with each other through heterogeneous networks using, for example, wireless Local/Personal Area Networks (WLAN/WPAN), cellular, wired Ethernet or Power Line Communications (PLC).

Here we present an industry perspective and overview of various standardization activities of M2M communications. This overview is by no means exhaustive: there are a plethora of standards and standardization activities in this area: indeed, this wealth of activity is arguably one of the challenges of this subject. We analyze in particular some of the enabling technologies, some of the key standardization efforts (particularly from our European perspective), and, due to lack of space, focus on applications of M2M to the industry sectors of Smart Grid and e-Health as two significant current areas.

The rest of the paper is organized as follows. Section 2 briefly summarizes some of the key challenges of M2M. Section 3 discusses several generic M2M standards. Sections 4 and 5 focus, respectively, on M2M standards for e-Health and Smart Grid. Section 6 describes research issues and directions and Section 7 concludes the paper.

## II. KEY CHALLENGES OF M2M

The main objective of M2M system design is to make sure that the M2M communication network is reliable and scalable and has the ability to support new functionalities to enable new business models as well as cost savings. Key challenges:

- *Lots of devices and data*: M2M communication networks could potentially be very large, supporting increasing numbers of deployed devices. Device capabilities can vary in terms of energy (battery or mains), processing and communications. There will be an explosion of data collected in such networks, with the emerging "big data" challenge of how to interpret and make good use of these data.
- *Quality of service and reliability*: Different applications have different Quality of Service (QoS) and reliability requirements that potentially need to be supported in an M2M network. For example, latency and bandwidth requirements for e-Health related traffic could be different from those for routine smart-metering; some traffic may need to be prioritized over other non-critical traffic.
- *Efficient networks and protocols*: Network protocols will have to deal with inherent characteristics of M2M networks such as long sleep cycles, energy and processing power constraints, time-varying radio propagation environments, and topologies varying with node mobility. To this end, energy efficiency is very important in M2M protocol design.
- *Security*: For large scale M2M deployments, security must be taken seriously from the outset and a holistic approach is recommended. Authentication, authorization, access control, and key management are some of the necessary components of a secure M2M system.

These general requirements may not apply in all scenarios and applications. Telemetry-based e-health Body Area Networks (BANs) with a few battery-powered devices differ massively from Home Area Networks (HANs) with large numbers of mains-powered devices.

Common standards (e.g. oneM2M below) must therefore consider these differing requirements. Configurable (cf. traffic classes in IEEE 802.11e) or modular (cf. Bluetooth profiles) approaches may be preferable to a "one-size-fits-all" approach.

## III. M2M STANDARDS

The M2M landscape has seen multiple standardization efforts in recent years. Whilst the commonality of purpose, direction and enthusiasm is welcome, this brings its own challenges with numerous standards vying for attention in the industry. Examples include efforts being made by groups such as the European Telecommunications Standards Institute (ETSI), Third Generation Partnership Project (3GPP) and the Internet Engineering Task Force (IETF). There are many more, but, due to lack of space, we have to focus on a select few, in what we hope is a representative selection of relevant standards. In all cases, the pull from industry is strong, needing a "speedy but not hasty" development of standards to support interoperation, global markets and economies of scale, but standardization must not slow the growth of this area.

One significant effort to rationalize and align this explosion of M2M standards is the oneM2M Partnership, formed in late 2012 by global several Standards Development Organizations (SDOs). The remit of this group is to address the significant risk of fragmentation and duplication of work among the SDOs by developing a common M2M Service Layer. The group is evaluating and merging all SDOs' standards into the baseline specifications of a common solution.

### A. ETSI Technical Committee (TC) M2M

ETSI TC M2M is focused on [3] service and operational requirements for M2M solutions and an end-to-end architecture for M2M, by identifying and filling standardization gaps, such as interconnection and integration of wireless capillary networks and their devices with wide area telecoms networks. The group collaborates closely with ETSI's activities on Next Generation Networks and also with 3GPP (below). As oneM2M (above) gathers pace, ETSI M2M will focus on capturing the EU specific requirements and responding to EU regulatory purposes (including mandates).

The general architecture of M2M networks such as that being specified in ETSI TC M2M is shown in Figure 1 (based on information and a similar diagram in [3]) with definitions given in Table 1.

Table 1: Key M2M Elements

| M2M Element | Role |
|---|---|
| M2M Area Network (Device Domain) | M2M Device/Gateway connectivity |
| M2M Gateway | Interworking and interconnection to the wider communication network |
| M2M Communication Network (Network Domain) | Communications between the M2M Gateway(s) and M2M Application(s) |
| M2M Applications | Contain middleware to support application services |

In Figure 1, M2M Devices (intelligent and communication enabled) form an M2M Area Network, perhaps a small-scale home environment or a much larger factory environment. The Area Network can employ various WLAN/WPAN technologies. The Access Network, connecting the Gateway to the Core can employ various Wide/Local Area Network (WAN/LAN) technologies. The M2M Gateway at the edge of the Device Domain connects to the M2M Management Server on the M2M Service Platform, subsequently reaching the M2M Applications.

ETSI M2M defines several use-cases: Smart Grid, connected consumer, city automation, automotive systems, and e-Health. Figure 2 [3] depicts ETSI's high level system overview of M2M. M2M defines three interfaces and models these interfaces following a RESTful [4] approach (RESTful indicates compliance with REpresentational State Transfer architectural constraints, see below). Interface *mIa* is the interface between application and service capabilities in the M2M Core. Service capabilities are exposed via this interface to applications as service capability features. *dIa* is the interface between application and service capabilities in the M2M Device / Gateway. *mId* is the interface between service capabilities in the M2M Core and the M2M Device / Gateway.

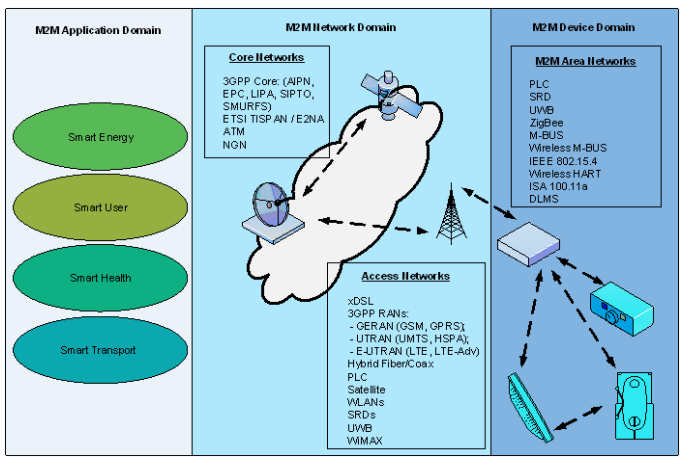

Figure 1: M2M Communication Network Architecture

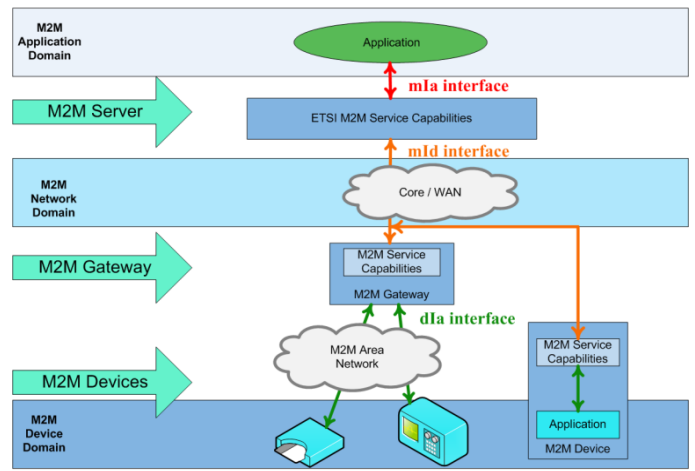

Figure 2: M2M High-level System Overview

## B. 3GPP Machine Type Communications (MTC)

3GPP has also been very active in M2M technology. 3GPP's Machine Type Communications (MTC) considers the optimization of access and core network infrastructure, allowing efficient delivery of M2M services. The focus has been on technical issues such as MTC congestion and overload control, as well as low cost MTC devices [5]. Conventional cellular systems are designed mainly for human-to-human (H2H) traffic, while a huge number of new M2M devices will pose serious challenges to the Radio Access Network (RAN) and core network by generating a large amount of control and data traffic. A number of proposals have been proposed to address the overload issue, e.g. back-off adjustment, access class barring, and MTC prioritization [5]. Work on low-cost LTE MTC focuses on three main areas (radio, baseband, and operation), with discussions of techniques such as reduction of maximum bandwidth, single receiver RF chain, and reduction of transmit power.

## C. IETF Constrained RESTful Environment (CoRE)

Technologies such as semantic web, social networks and RESTful services [4] can provide some powerful tools to realize the "web of things" based on M2M and IoT. The IETF CoRE (Constrained RESTful Environment) working group is developing embedded web services for M2M/IoT applications. Notably, the CoAP (Constrained Application Protocol) protocol is an application transfer protocol realizing important RESTful features for resource-constrained devices in M2M networks [6]. In future smart homes, M2M networks will be seamlessly integrated with other home management systems (e.g. home energy and communication/entertainment networks) and with the wider web (e.g. social networks). In this sense, human networks and machine (sensors and objects) networks may become intertwined.

## D. Weightless

The Weightless Special Interest Group (SIG) [7] is a new proprietary wireless standard being developed for M2M using cognitive radio techniques in TV white spaces (unused guard frequency bands where devices can operate as secondary devices subject to strict interference requirements, without causing interference to the coexisting primary network). Weightless is a lightweight protocol that connects a base station with a large number of M2M devices, accessing the spectrum white spaces opportunistically. Nodes (machines) communicate with a base station depending on their locations and spectrum availability. Multi-hop relays are also deployed to improve coverage.

Control mechanisms and enablers for white space access are under consideration by regulators (e.g. the US's Federal Communications Commission, FCC; the UK's Office of Communications, Ofcom). Supporting technologies and standards are being developed where necessary by groups such as IETF's Protocol to Access White Space database (PAWS) working group.

## E. Open Mobile Alliance (OMA)

OMA has brought together a collection of industry forums including the WAP Forum and the Sync ML Initiative, resulting in a coherent approach for mobile service enablers. The OMA specifications are independent of the specific underlying cellular technologies carrying the traffic.

The OMA are already working in the M2M application area and are liaising with both 3GPP and oneM2M. They are standardizing the OMA Lightweight M2M enabler specification, which goes beyond traditional cellular to cover wireless sensor devices. The standard is secure (using public key methods) and has a very low overhead.

## F. ECHONET-Lite

The Great East Japan earthquake of March 2011 has had serious and ongoing repercussions for the Japanese energy industry. ECHONET-Lite is one of their responses: a Plug-and-Play, object orientated, wired/wireless technology agnostic open standard for Layers 5–7 of the ISO OSI stack, internationally standardized through ISO and IEC. It can interconnect everything in the home including energy, security and healthcare, linking it all securely to the wider Internet.

## IV. M2M STANDARDS FOR e-HEALTH

The 2011 International Telecommunication Union-Telecommunications (ITU-T) Technology Watch report notes that trends in e-Health point towards increasing focus on personalized medicine, standardized electronic health records, remote/tele- healthcare and diagnostics. A subsequent 2012 report emphasizes that e-Health standardization related to electronic health records is likely to be the key enablers of digital collection, storage and processing of medical data that will drive the e-Health revolution. Furthermore, there is the strong commercial motivation of a huge potential market, ranging from basic monitoring applications (such as blood pressure/heart rate monitors) through to therapeutic applications (such as active medical implants in the body).

Estimates highlighted in [9] from the European Society of Cardiology suggest that twelve million people may suffer from a heart failure condition in Europe with forty percent of these patients likely to require active implantable medical devices. This translates to roughly five million implants by 2015. Similarly, there is also the increase in the number of diabetic patients. Currently there are 32 million diabetic patients in Europe using neuro devices, and this market is growing rapidly (€2 billion per year from current trends) [9].

Therefore, e-Health is a key application of M2M currently under active consideration in different standards fora such as ETSI TC M2M, ETSI TC e-Health, ITU-T Focus Group (FG) on M2M etc. According to [8], M2M applications for e-Health enable the remote monitoring of patient health and fitness information, the triggering of alarms when critical conditions are detected, and in some cases also the remote control of certain medical treatments or parameters. A sensor BAN is typically deployed around the patient to record health and fitness indicators such as blood pressure, body temperature, heart rate and weight. These sensors often have form factor and power constraints, and are connected to the remote control center via a gateway or data aggregator over some short range wireless technologies. ETSI M2M defines [8] a number of e-Health applications: remote disease management, aging independently, and personal fitness and health improvement. Compared to the M2M applications in support of disease management and ageing independently, the support of personal fitness and health improvements will most likely require less frequent logging or uploading of data and is more tolerant to delays.

In certain patient and medical implant monitoring applications, large volumes of data will be transferred to the monitoring station/application. There is a need for identifying new radio spectrum for the devices to operate in to facilitate such monitoring [9]. Additionally, these devices should not cause interference to existing medical implant devices and should co-exist gracefully with other communication services (if any) operating in the same spectrum. The need for harmonizing such spectrum across the whole of Europe and ideally, worldwide has been highlighted [9]. There appears to be a general consensus that the over-crowded Industrial Scientific and Medical (ISM) band is not suitable for the aforementioned applications. It has also been identified that the power consumption of devices operating in these bands do not meet the power constraints imposed on medical implants (e.g. having enough battery to last for at least ten years). In [9], the Electronic Communications Committee of the European Conference of Postal and Telecommunications Administrations (CEPT ECC) was requested to identify 20MHz of spectrum in the frequency band 2300-3400MHz for the above application. In response, the CEPT ECC identified the frequency band 2483.5-2500MHz as the most suitable for this application. On the other hand, the US FCC has approved the use of 2360-2390MHz band for applications in indoor healthcare environments and 2390-2400MHz band elsewhere (e.g. in hospitals, in home, ambulances, etc.)

Additionally, new proposals within ETSI TC M2M and ETSI TC e-Health identify needs in: BAN – monitoring, preventive, therapeutic, fitness, and well-being applications; Telemedicine – identification of use case scenarios, profiles and appropriate protocols to facilitate this; and e-Health network architecture and protocols to support BAN and telemedicine applications scenarios. However, at the time of writing, not much progress has been made on these proposals.

At the same time, the ITU-T M2M focus group is also trying to identify requirements for M2M standardization, looking at existing standards with the aim of identifying whether they meet these requirements. The intention is to identify work items based on gap analysis, with the findings from this study yet to emerge. As highlighted in [10], the actual standards work to be undertaken is likely to be decided based on the outcome of this study. This report also highlights several barriers to standardization, some of the prominent ones being the existence of proprietary systems, the massive amounts of data being collected from these systems; the lack of standard content format and the lack of open freely available standards.

## V. M2M STANDARDS FOR SMART GRID

### A. Background on Smart Grids

Smart Grid is the deployment of a next generation power infrastructure able to support intermittent renewable energy supplies, bidirectional flows from prosumers (customers that both produce and consume energy) and increasingly intelligent management of assets. Smart-metering, an important element of Smart Grid, is the mass-deployment (e.g. to all UK homes and small/medium enterprises by 2019: 53 million meters) of replacement electricity, gas, water and (where applicable) heating meters that enable fine-grained meter reading, variable tariffs, instantaneous consumer display of energy usage and the ability for prosumers to change energy companies readily. The first phase of Smart Meters will provide some initial basic support for future evolutions to incorporate demand side management of local appliances via home energy management systems, the management of prosumer-side energy caching and storage facilities, and the export of energy produced by micro generation.

This vast undertaking has necessitated continent-wide international cooperation from the outset across Europe. The European Union (via the European Commission executive) issued three mandates (M/441: Smart Meters; M/490: Smart Grid; M/468: Electric Vehicles), granting standards bodies remits to develop the required standards frameworks. To respond to these mandates, the three main European SDOs formed joint working parties to coordinate efforts and tasks, such as the Smart Metering Coordination Group (SM-CG) and the Smart Grid Coordination Group (SG-CG). Each SDO has been allocated specific standards development work according to expertise: ETSI M2M for inter-device communications, CENELEC (European Committee for Electrotechnical Standardization) for the next generation of electricity meters, and CEN (European Committee for Standardization) for the next generation of non-electricity meters. Several official reports and responses (e.g. [11]) have been presented to the European Commission, which are contributed to, reviewed and commented on by individual nation's national standards bodies (e.g. the UK's British Standards Institution, BSI).

We now introduce several Smart Grid related standards mainly from an M2M perspective in two areas: smart-metering and home energy management. Smart Grid standards such as IEEE P2030, NIST and ANSI are not covered here and the interested reader is referred to [12].

### B. Global Smart Metering Standards

**ETSI TC M2M**

ETSI TC M2M has recently approved a document [13] which discusses several detailed use cases in relation to a typical smart-metering configuration. Examples of these use cases include: obtain meter reading data, install, configure and maintain the smart-metering information system, support prepayment functionality, monitor power quality data, manage outage data, etc. All of these use cases are discussed in terms of general description, stakeholders, scenario, information exchanges, and potential new requirements.

A number of liaisons have been established with other standardization bodies, e.g. CEN, CENELEC and the ZigBee Alliance. TC M2M's responsibilities to M/441 include: providing access to the meter databases through the best network infrastructure (cellular or fixed); providing end-to-end services capabilities, with three targets: the end device (Smart Meter), the concentrator/gateway and the service platform. Further, smart-metering application profiles will be specified including service functionalities.

**DLMS/COSEM**

The Meter–to–M2M-Gateway interface is based on the IEC 62056 set of standards (otherwise known as the Device Language Message Specification/Companion Specification for Energy Metering, DLMS/COSEM) protocol, defining data exchange protocols for meter reading, tariff and load control.

The IEC 62056 series is the International Standard version of the DLMS User Association (UA)'s four specification documents called the Green, Yellow, Blue and White Books (meter object model, architecture and protocols, conformance testing and glossary respectively). If a product conforms to the Yellow Book, it complies with IEC 62056.

**M-Bus**

M-Bus ("Meter-Bus") is a European standard (EN 13757-2 physical and link layer; EN 13757-3 application layer) for remote reading of meters, as well as various sensors and actuators. The M-Bus concept was based on the ISO-OSI Reference Model, to implement an open system which could utilize almost any desired protocol, but (since it is not a network) omits certain unnecessary layers, providing only the Physical, Data Link, Network and Application layers.

M-Bus was developed to provide a system for the networking, and remote reading, of utility meters. M-Bus fulfils the special requirements of remotely powered or battery driven systems, including consumer utility meters. When interrogated, the meters deliver the data they have collected to a common master, such as a hand-held computer, connected at periodic intervals to read all utility meters of a building. An alternative method of collecting data centrally is to transmit meter readings via a modem.

Alarm systems, flexible illumination installations, and heating control are other M2M-type applications for M-Bus.

*C. HEMS (Home Energy Mangement System)*

A number of different technologies present themselves as enablers for a future HEMS, a major application area of M2M in homes, offering the ability for domestic appliances to communicate with each other and with the Smart Meter or HEMS hub. Two frontrunners are wireless low power radios, and wired connections over existing mains wiring.

**IEEE 802.15.4, ZigBee and ISA 100.11a**

The IEEE 802.15 standards began with the adoption and adaptation of the Bluetooth specification, subsequently extending the standard in a number of directions, including mesh networking (802.15.5), BANs (802.15.6, above) and for low-power applications such as sensor networks (802.15.4).

IEEE 802.15.4 is a low data rate (250kbps) variant operating in the ISM bands with explicit support for low power modes of operation. The ZigBee Alliance has produced a suite of specifications sitting above the 802.15.4 MAC, defining higher layer interoperability for *ad hoc* mesh networking applications.

Further systems have also adopted 802.15.4 as their radio technology, such as IETF's 6LoWPAN, WirelessHART and ISA 100.11a. The last two are from an industrial automation process control background. ISA 100.11a introduces some interesting additions that improve the dependability of M2M networks, including a configurable TDMA air interface where time slots can be used in a number of ways (devices can be given dedicated time slots for predictable, regular traffic; alternatively, multiple devices may be given a shared, prioritized CSMA-CA contested time slot for infrequent events such as alarms and bursty traffic).

**PLC**

PLC is an appealing technology for HEMS, modulating data signals along the existing mains power wiring (cf. ADSL over telephone wires). HEMS-controlled appliances are already plugged into the mains, so leveraging that existing connection seems a "neat" solution. Some construction materials may make wireless impractical (e.g. some old housing stock in Europe with walls largely impenetrable to wireless signals). However, there are limitations: networks are limited to the same electric circuit (i.e. struggle to jump across fuse boxes to other circuits) and channel conditions are noisy and far from ideal (even in comparison to wireless channels).

Products already exist in the market place with a number of mutually incompatible proprietary solutions (e.g. HomePlug, www.homeplug.org; HomeGrid, www.homegridforum.org).

**IEEE P1905.1 Home Networking Convergence**

IEEE P1905.1, now approved by the IEEE Standards Association, converges disparate technologies (not just PLC, but also wireless, Ethernet and coaxial) into a hybrid home network standard, by defining an abstraction layer to support the interworking of complementary technologies and composite solutions. P1905 has had some issues when component solutions cannot interoperate and actually interfere with each other (e.g. HomePlug vs. HomeGrid). The next steps for P1905.1 is to identify and integrate yet more underlying transport technologies.

## VI. RESEARCH ISSUES AND DIRECTIONS

*A. Wireless Technologies for M2M Area Networks*

This is an active research area with no clear winner so far. In addition to conventional WPAN/WLAN technologies (e.g. ZigBee, Bluetooth, Wi-Fi), M2M wireless networks based on cognitive radio technology have also been proposed. For example, secondary use of the 2360–2400MHz band for medical BANs will enable more flexible and efficient use of spectrum. Intelligent antennas and multi-hop connections are also promising for realizing reliable M2M networks. For Smart Grid, the end-to-end performance of heterogeneous networks covering multiple network domains is an open question.

*B. Security and Privacy*

M2M security focuses on several attributes of a user and their communications, including *authenticity* (the user is who they claim); *authority* (the user is allowed to perform the operation requested); *integrity* (the data received is the same as that transmitted); and *confidentiality* (communication between users remains secret if intercepted) [14].

M2M privacy considers the relationship of data to users and the needs of M2M application professionals. For instance, e-Health patient users have a right to privacy of their data, but it is essential that authorized health professionals have access to their patients' health records. Also, smart-metering privacy is a well-known problem and has received much attention recently.

Standards have also been active in defining security and privacy mechanisms for effective deployment of autonomous and largely unassisted M2M applications as well as addressing customer confidence. To this end, ETSI M2M has incorporated measures such as Protocol for Carrying Authentication for Network Access (PANA) [15] and Advanced Meter Sign On (AMSO) to securely transport authentication information and other bootstrapping-related attributes.

*C. Data Analytics for M2M*

Entities and smart objects in an M2M network discover and explore the environment and adapt to enhance system capability and reliability. Further, self-configuration learning is essential when an M2M system encounters a new environment with interference and dynamic changes. Making informed decisions and extracting intelligence from the data in M2M communications is non-trivial. The Smart Grid, for example, can be seen as a huge sensor network, with immense amounts of grid sensor data from various sensors, meters, appliances and electrical vehicles. Data mining and predictive analytics are essential for efficient and optimized operation of the grid. Traditionally, power networks are protected and controlled using SCADA technology: very often the central control system is not fast and intelligent enough for modern power systems [16]. Future Smart Grid systems will deploy new monitoring devices such as Smart Meters and Phasor Measurement Units (PMUs) to enable a Wide Area Measurement System (WAMS), generating vast data every day. A key question is how to analyze and process these data in an efficient and timely manner. Various machine-learning techniques can be used here for data analysis and processing.

*D. New Network Architectures for M2M*

Cloud computing is a new computing paradigm that has significantly changed the global ICT landscape. Since M2M networks will generate vast amount of data and have many resource-constrained devices, it can be envisaged that cloud computing will play a key role in M2M development by offering desirable features such as mass data storage, data offloading, processing and virtualized infrastructure. For example, there has been recent work on applying publish/subscribe cloud concepts to distributed monitoring and control of Smart Grid networks. However, it remains to be seen what business models will be most suitable for M2M, e.g. Infrastructure as a Service (IaaS), or Software as a Service (SaaS), etc. Further, there is a trade-off in utilizing cloud service for M2M applications: communication latency (machines/cloud) and energy consumption vs. computation/processing overhead. As research in cloud computing and M2M evolves, we will see more and more M2M applications delivered and managed in the cloud.

VII. CONCLUSIONS

We have presented an overview of several standardization activities of M2M communications, focusing on two major applications of M2M: e-Health and Smart Grid/meters, and analyzed some of the enabling technologies and research challenges. It is difficult to directly compare such disparate and often complementary standards efforts, however, Table 2 offers an overview, and picks out key attributes and capabilities to highlight the differences in capability, scope and applicability:

Table 2: Overview and Comparison

| SDO | Scope | Properties |
|---|---|---|
| **End-to-End Systems, Middleware** | | |
| oneM2M | Global | (Service layer): Adopting the best architecture and solutions from around the world. |
| ETSI | Europe | **M2M** (service layer): handing architecture over to oneM2M. Then focusing on European mandates and variations |
| 3GPP | Global | **MTC**: Radio and core network. Focusing on optimizing cellular networks for M2M traffic. |
| IEEE | Global | **1905.1** brings together wired and wireless standards in the home. |
| Echonet | International standard, but primarily Japanese | **Echonet-Lite**—home networking for energy, security and healthcare. |
| **Enabling Components** | | |
| IETF | Global | **CoRE/CoAP**: embedded web services for M2M devices. |
| Weightless | Global | **Weightless**: cognitive radio white-space wireless standard. |
| OMA | Global | **Lightweight**: Application enablers for cellular, being applied to M2M and sensors. |
| IEC | Global | **DLMS/COSEM**: the meter-to-M2M Gateway interface. |
| EN | Europe | **M-Bus**: remote meter reading. |
| IEEE | Global | **802.15.4** Wireless PAN standard and its progeny: front runner for HEMS. |

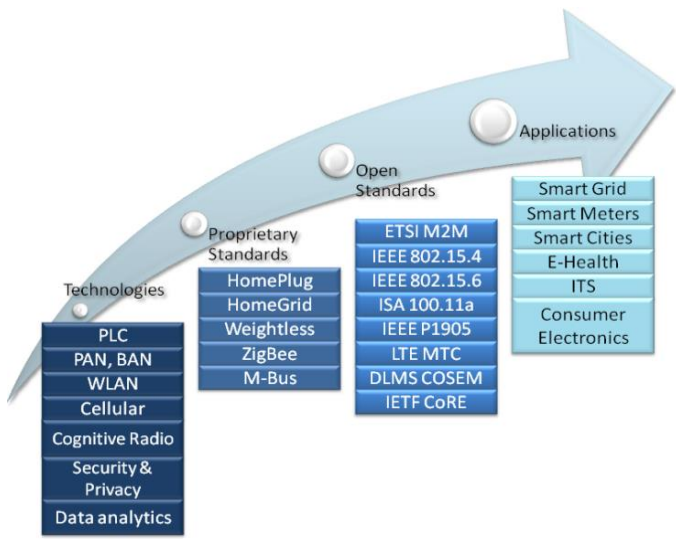

Figure 3: M2M Technologies, Standards and Applications

Figure 3 summarizes the M2M technologies and standards discussed in this paper, illustrating how they come together for the fast-approaching future.

The roadmap of worldwide M2M deployment is still not clear, but IoT empowered by advanced ICT technology will almost certainly be as life-changing as the current Internet. M2M communications underpin this vision, so we envisage that M2M will be an exciting research area for engineers in both industry and academia for many years to come.